\documentclass[11pt,a4paper]{article}
\usepackage{aas_macros}
\usepackage{graphicx}
\usepackage[table]{xcolor}
\usepackage[margin=2cm]{geometry}
\usepackage{hyperref}
\usepackage{titling}
\usepackage[square,numbers,sort&compress]{natbib}
\usepackage{enumitem}

\usepackage[noindentafter]{titlesec}
\begin{document}

\begin{titlepage}

\vspace*{6cm}

\centering

{\Huge UK White Paper on Magnetic Reconnection}
\vspace{1.5cm}

{\huge Submitted to UK Space Frontiers 2035}
\vspace{1.5cm}

{\Large Primary Thematic Area: Heliophysics}
\vspace{1.5cm}

{\Large 28 Nov 2025}
\vspace{1.5cm}

{\large \textbf{Lead Author and Point of Contact}\vspace{0.2cm} \\
Alexander J. B. Russell\vspace{0.1cm}\\
School of Mathematics \& Statistics, \\ University of St Andrews. \\
ar51@st-andrews.ac.uk}
\vspace{0.75cm}

{\large \textbf{Co-authors and Signatories}\vspace{0.2cm} \\
A list of authors and signatories is appended at the end.}\vspace{0.5cm}

\end{titlepage}

\titlespacing\section{0pt}{5pt plus 2pt minus 2pt}{4 pt plus 2pt minus 2pt}
\titlespacing\subsection{0pt}{5pt plus 2pt minus 2pt}{1pt plus 1pt minus 1pt}
\titlespacing\subsubsection{0pt}{3pt plus 1pt minus 1pt}{1pt plus 1pt minus 1pt}
\titlespacing\paragraph{0pt}{3pt plus 1pt minus 1pt}{5pt plus 1pt minus 1pt}

\section*{Executive Summary}
Magnetic reconnection powers explosive releases of magnetic energy, heating and particle acceleration throughout the plasma universe. Knowledge of this universal process is vital to understanding the Heliosphere, as it plays a key role in solar flares, coronal mass ejections, coronal heating, solar wind acceleration, geomagnetic storms, and interactions between the solar wind and planetary magnetospheres. As such, reconnection underpins multiple science objectives of multiple future space missions. The UK plays a leading role in this international field, through a combination of in situ measurements from Earth’s magnetosphere and the solar wind, observations of the solar corona and chromosphere, and world-class numerical simulations and theory. This white paper identifies:

\begin{itemize}[itemsep=-3pt,topsep=0pt]
    \item {Nine priority science objectives for reconnection research in the next decade;}
    \item {Recommendations to guide investment in theory, simulations and infrastructure;}
    \item{Mission priorities and required measurements to ensure the UK maintains and improves its world-class credentials in reconnection science.}
 
\end{itemize}


\section{Introduction: Magnetic Reconnection and Its Impacts}\label{sec:into}

Magnetic reconnection is a fundamental plasma phenomenon in which spatially-localised changes in magnetic field connectivity change the system on global scales, typically accompanied by explosive release of magnetic energy, high-speed outflows, heating, and particle acceleration. 

In the Heliosphere, magnetic reconnection powers solar flares, coronal mass ejections (CMEs) and geomagnetic storms and substorms. It accelerates particles to very high energies contributing to radiation hazards, self-organises magnetic fields, mediates mass and energy transport between the solar wind and planetary magnetospheres, and has a central role in turbulence. It is also believed to play an important role heating the solar corona and accelerating the solar wind.
Astrophysics applications include stellar flares, stellar CMEs and exoplanet magnetospheres, which all affect habitability. Furthermore, reconnection's influence on instabilities, turbulence and transport makes it an important aspect of star formation, accretion disk dynamics, planet formation, active galactic nuclei, interstellar medium turbulence, pulsar magnetospheres, magnetar flares and gamma ray bursts \citep{2011Uzdensky}. Reconnection is also important in lab plasmas, where controlling it is central to advancing fusion.

Heliophysics has been a primary driver of advances in reconnection since its inception. The Heliosphere provides a virtual lab where reconnection can be studied across a far broader range of scales than can be achieved in experiments. It also provides a wealth of in situ and remote observations at resolutions that are rarely achievable for astrophysical objects. At the same time, the heliospheric reconnection community have an excellent track record of collaborating with experimentalists and astrophysicists to drive scientific progress across diverse applications.

Planning and prioritisation exercises globally have for many decades recognised reconnection as one of the most important science drivers in astronomy, solar physics and space physics, reflecting its importance as a fundamental process that underpins space weather and many of the most important science questions about the Sun, Solar System and wider Universe. This international recognition makes reconnection a key opportunity for international partnerships and UK investment.


\section{Strategic Context}\label{sec:strategic-context}

Reconnection powers the solar flares and geomagnetic storms responsible for severe space weather, which is assessed 4-4 in the UK National Risk Register \citep{2025NRR}.
Controlling reconnection is central to advancing fusion energy, with UK Government having committed over £2.5 billion to create a commercial reactor by 2040 as part of the UK Industrial Strategy \citep{2025IndustrialStrategy}. UK leadership in instrumentation attracts multi-million pound contracts, e.g. ESA contract for the SOLAR-C EUVST short-wavelength camera (details in the white paper by Matthews). Reconnection is also key to new technologies, e.g. UK-funded Magnetic Reconnection Plasma Thrusters for Spacecraft Propulsion (Southampton).

Funding reconnection science directly invests in the UK's current and future workforce -- training individuals who are prized by industry for their coding, supercomputing, machine-learning and AI, data science, 
mathematical and analytical skills. Besides industry, many \lq{reconnection alumni}\rq{} go on to work in education, science communication, policy, government and defence, while others remain in science in the UK or overseas, where they contribute to scientific discovery and UK soft power.


\section{Scientific Motivation}\label{sec:motive}

Advances in reconnection rely on five research pillars, with the UK having a world-class record in observations, simulations and theory, and an exceptional talent for joining them together. This section showcases some major achievements from the last 15 years, which have substantially addressed priority questions from previous decades. These highlights provide context for the objectives in Section \ref{sec:obj}, and demonstrate the importance of strategically funding all five pillars and their synergies.

\subsection*{[Pillar 1]  Direct In Situ Measurements}

\textbf{Magnetospheric Multiscale (MMS),} launched in 2015, has been a defining mission for collisionless reconnection. Four spacecraft in a tetrahedral formation resolved the electron diffusion region for the first time at the dayside magnetopause \citep{2016Burch} and nightside magnetotail \citep{2018Torbert}. The reconnection-focused mission has produced a huge scientific return, e.g. review \citep{2020HesseCassak}, including confirming predictions of collisionless reconnection theory/simulations, discovering \lq{electron only}\rq{} reconnection \citep{2018Phan} and characterising turbulence-driven reconnection in Earth's magnetosheath \citep{2022Stawarz}.

\textbf{Parker Solar Probe (PSP)} has approached within a record 9~$R_{\odot}$ (0.04~AU) of the Sun, encountering reconnection in the heliospheric plasma sheet, post-CME plasma sheets \citep{2025Patel} and magnetic switchbacks \citep{2021Froment}.
PSP and Solar Orbiter encounter ejecta from solar reconnection unprecedentedly close to the source, e.g. \citep{2025Desai}, significantly adding to \lq{system view}\rq{} constraints on coronal reconnection.

\textbf{Energy Partitioning} In situ data have unlocked empirical heating relations \citep{2009Drake,2013Phan,2014Phan,2017Misty,2020Tilquin}, 
providing compelling evidence that reconnection preferentially heats ions with $\Delta T_p/\Delta T_e \approx (m_p/m_e)^{1/4} \approx 6.5 $ \citep{2018Hoshino,2025Oka}. UK-led investigations of energy transport for collisionless reconnection have shown that ion enthaply flux is dominant in the exhaust \citep{2013Eastwood} but electron enthalpy flux dominates in the electron diffusion region \citep{2020Eastwood, 2024Fargette}. Building on the in situ evidence and spectroscopic observations of solar flares, a UK-led team this year proposed that \lq{hot ions}\rq{} are also likely in flares \citep{2025Russell}. 

\subsection*{[Pillar 2]  Remote Observations}

\textbf{System View} Solar observatories have been successfully coordinated to capture the system response to reconnection, including outflow jets \citep{2013Imada,2014Tian}, termination shocks \citep{2015Chen,2018Polito,2024French}, heating \citep{2011Fletcher}, turbulence \citep{2017Kontar} and nonthermal particles \citep{2010Krucker,2014Krucker,2022Fleishman,2018Omodei}. The continuous full disk EUV imaging of SDO AIA has provided an essential foundation, augmented by EUV spectroscopy (EIS, IRIS, SPICE), X-rays (RHESSI, STIX, NuSTAR), gamma rays (Fermi) and radio (e.g. LOFAR). The X8.2 SOL2017-09-10 flare has been exceptionally well studied, with imaging capturing the flare plasma sheet, spectroscopy constraining ion temperatures and turbulence \citep{2018Warren,2018Cheng,2018Li} and evidencing nonthermal ions \citep{2018PolitoKappa}, while radio, X-rays, and gamma rays have revealed particles with energies up to 300~MeV \citep{2022Fleishman,2018Omodei}. 

\textbf{Hi-Res Imaging and Spectroscopy}\label{sec:motive:hi-res}
IRIS's 0$^{\prime\prime}$.33 spectrograph has achieved the breakthrough of regularly resolving hot chromospheric evaporations -- a key capability for constraining energy fluxes from reconnection \citep{2014Tian,2015Graham,2015Polito,2015Young}. IRIS's Slit-Jaw Imager has discovered nanojets \citep{2021Antolin} and flare ribbon features consistent with tearing of the flare current sheet \citep{2021WyperPontin,2021French}, while Solar Orbiter's HRI has obtained the most detailed EUV coronal images, documenting explosive and persistent reconnection at 3D null points \citep{2023Cheng}, oscillatory reconnection \citep{2024Kumar} and relaxation of magnetic braids \citep{2022Chitta}.

\textbf{Solar Magnetic Field Data}
Continuous full disk $1^{\prime\prime}$ magnetograms from SDO HMI, combined with coronal magnetic field modelling and data-driven simulations, have provided key information on the solar magnetic field evolution that triggers flares/eruption \citep{2014Amari,2021YardleyFormation,2021YardleyStealth}, and 3D topologies in which reconnection occurs, from single 3D nulls \citep{2017Wyper,2018Wyper} to active region networks of $\sim$100 null points \citep{2025PesceRollins}.

\subsection*{[Pillar 3]  High-Performance Simulations}

\textbf{Self-generated Turbulence} has been captured in reconnection layers by 3D PIC \citep{2011Daughton} and 3D MHD simulations since 2011 and 2015 \citep{2011Daughton,2015Oishi,2016HuangBhattacharjee,2017Beresnyak}, which has ushered in a new era for reconnection modelling. With moderate or strong guide fields, self-generated turbulent reconnection rates of 0.01 are found in 3D MHD \citep{2016HuangBhattacharjee,2022Beg} and 0.1 with collisionless physics \citep{2011Daughton,2014Daughton,2024Huang}. MHD rates $\geq$0.03 have been found in the absence of a guide field \citep{2026Ming}, and when transient turbulent forcing is applied at early times \citep{2025Vicentin}. 

\textbf{Particle Energisation} 3D PIC simulations are now large enough to capture heating and acceleration of electrons and ions, e.g. \citep{2021Zhang}. Like observations, simulations indicate ions are preferentially energised, potentially starting a major shift away from a previous focus on electrons.
New approaches that achieve greater scale separations are being developed, e.g. kglobal \citep{2021Arnold} and DISPATCH \citep{2025Haahr}.

\textbf{Model in the Mission}
A major recent trend is integration of simulations and theory into missions for development and interpretation. PIC simulations and theory for MMS were funded as specific mission packages \citep{AboutMMS}. 1D radiative hydrodynamics and forward-modelled MHD simulations have amplified the success of IRIS. ESA funding to develop back mapping tools has been crucial to the success of Solar Orbiter. In the UK, coupling science simulations with instrument response models has greatly informed the development of SMILE SXI (Leicester) and SOLAR-C EUVST (MSSL).

\subsection*{[Pillar 4]  Theory}

\textbf{Reconnection Rates} Today, multiple theories produce fast rates of 0.01 to 0.1, including Petschek, collisionless X-line, plasmoid-mediated \citep{2009Bhattacharjee} and turbulent reconnection \citep{1999Lazarian}. Recent UK work has shown how plasmoid-mediated principles can extend to 3D despite the presence of turbulence \citep{2025Russellb}.

\textbf{Tearing Instability} There is now a strong consensus that long current sheets fragment -- a major change from the classic picture of a single X-line. It is also widely accepted that \lq{ideal tearing}\rq{} instabilities grow on system time scales for current layer thicknesses far exceeding the Sweet-Parker ratio \citep{2007Loureiro,2014PucciVelli}, which provides one possible solution to the triggering problem, although other factors such as loss of equilibrium and reconnection-rate dependencies provide other possible solutions.

\textbf{3D} Oblique tearing modes and kink instabilities break translational symmetries, introduce field line ergodicity and greatly alter reconnection and particle acceleration when comparing 3D and 2D. Theory for describing the 3D topological skeletons is now mature and regularly applied to eruptions involving single 3D nulls. Recent UK-led work has highlighted the importance of magnetic separators  as reconnection sites \citep{2024Parnell}, and separator networks linking around 100 null points are now being connected to flare observations \citep{2025PesceRollins}, moving solar applications beyond single 3D nulls and quasi-separatrix layers.

\subsection*{[Pillar 5]  Experiments}
While lab experiments are outwith the main scope of the Space Frontiers exercise, it must be remarked that they are a fifth pillar of reconnection research. Recent highlights include validation of collisionless and collisional reconnection regimes and plasmoid instabilities \citep{2022Ji,2023Ji}. The new FLARE facility in the USA will probe larger scale separations and Lundquist numbers \citep{FLARE}. Traditional experimental setups are also complemented by reconnection studies in laser plasmas, e.g. \citep{2018Kuramitsu}.


\section{Science Objectives for 2026--2035}\label{sec:obj}
The following objectives were selected following community input via an online survey and two online discussions. The objectives that emerged are aligned with recent international reviews \citep{2022Ji,2025Nakamura}, US Decadal Survey white papers \citep{2023Decadal} and the 2022 STFC SSAP Roadmap, but reflect current UK expertise and priorities. The following objectives are very interlinked and therefore explicitly not ranked. 

\subsection*{[Objective 1] Cross-scale Coupling Between Kinetic and Fluid Scales}
Now that MMS, simulations and experiments have revealed the kinetic physics of collisionless reconnection, the obvious next goal is cross-scale coupling between kinetic and fluid scales. This is also critical to understanding nonthermal particles in solar flares. 
How does the microphysics influence the macrophysics and vice versa? Can we incorporate the microphysics into fluid theories (closure problem)? Addressing this challenge requires simultaneous in situ measurements at kinetic and fluid scales, which Plasma Observatory directly addresses.  Kinetic simulations with large domains are vital, requiring investment in world-leading HPC, PIC expertise and novel code frameworks that self-consistently bridge the kinetic and fluid scales, e.g. following kglobal \citep{2021Arnold} and DISPATCH \citep{2025Haahr}.

\subsection*{[Objective 2] Cross-scale Coupling at Fluid Scales}
In large systems like the solar corona, reconnection requires cross-scale coupling across many orders of magnitude of fluid scales. Plasmoid instabilities and reconnection-driven turbulence have emerged as front-runners to explain this coupling, with both able to produce fast reconnection rates without needing kinetic effects. However, questions remain about how cross-scale coupling works in 3D. When do turbulence or plasmoids dominate? What are the properties of reconnection-driven turbulence? How does small scale structure such as fragmentary 3D current sheets develop and evolve? How is fine scale structure in flare ribbons and auroras related to reconnection and instabilities? How does turbulent reconnection extend to 3D nulls and separators? 3D MHD simulations, forward modelling, solar observations and theory are all crucial tools to address these questions.

\subsection*{[Objective 3] Energetics and System View}
Advances on energisation at the reconnection site itself now open up larger questions about energy conversion and partitioning throughout the global system. The full set of relevant processes includes Alfv\'enic jets, shocks, turbulence, waves and particle acceleration. Are acceleration and heating multi-step? How is energy shared between electrons and ions and what governs this partitioning? What are the contributions to heating and acceleration in the electron and ion diffusion regions, at separatrices and downstream? What constitutes the energy flux away from the reconnection site? Is there footpoint acceleration in flares like in auroras? Tackling these questions requires 3D modelling of large systems with realistic boundary conditions, e.g. building on \citep{2021Zhang,2022Shen}.
Integrating MHD with test-particles has proved valuable for modelling flares \citep{2020Gordovskyy} but more self-consistent approaches are needed in future.

\subsection*{[Objective 4] Reconnection Regimes}
Reconnection covers a large family of processes that occur under different conditions. This diversity is represented in the 2D reconnection phase diagram \citep{2011JiDaughton}, in which the system size, resistivity and ion inertial length control which of (at least) five types of reconnection occurs.
The number of  regimes is greatly expanded in 3D, which supports turbulent reconnection, finite-B (slipping) reconnection and reconnection at 3D nulls and separators. Simulations indicate that reconnection is more sensitive to the guide field in 3D than in 2D, how does this extend over the parameter space? Finite-B (slipping) reconnection and reconnection at 3D nulls and separators require different theoretical tools than current sheet reconnection, and this is an active area of research \citep{2025MacTaggart,2026Stanish}. There is also much still to learn about reconnection in partially-ionised plasmas, e.g. in the chromosphere.

\subsection*{[Objective 5] 3D Topologies}
The questions about cross-scale coupling, energetics and onset should explicitly be explored for reconnection at 3D null points and separators. Now that theory of 3D magnetic skeletons is established, simulations of eruptions involving isolated 3D null points are mature \citep{2017Wyper,2018Wyper}, and hi-res observations of coronal 3D nulls are becoming available \citep{2023Cheng}, we must aim to link simulations, observations and theory by modelling intensity, spectroscopic and particle signatures for 3D null point and separator reconnection. The coming decade must also move beyond single null points to address complicated topologies, such as the separator network connecting $\sim$100 3D null points studied by \citep{2025PesceRollins}, for which the UK contributed the topological analysis. How common are large separator networks? How does reconnection spread within them? Are emission hot spots often identified with spine field lines?

\subsection*{[Objective 6] Onset of Fast Reconnection}
CMEs lift off gradually before suddenly accelerating, and the flare impulsive phase (the main energy release, when many electrons are accelerated) is preceded by a preflare phase of about ten minutes (radio emission and $T_e\sim\ $10~MK ). Does the sudden change reflect a change in reconnection regime as parameters such as the guide field evolve, or a switch-on via instability or turbulence \citep{2025Stanish}? Does the guide field control electron acceleration \citep{2021Arnold}? Coronal heating by braiding also requires that finite-B reconnection switches on after energy has built up; what acts as the switch in that context? 

\subsection*{[Objective 7] Oscillatory and Time-dependent Reconnection}
The canonical reconnection theories are steady-state (Sweet-Parker and Petschek) or quasi-steady state with time-independent mean fields plus fluctuations (plasmoid-mediated and turbulent). How does reconnection differ when time-dependence is taken into account, either for short-lived bursts or oscillatory reconnection \cite{2009A&A...493..227M,2017ApJ...844....2T,2022MNRAS.513.5224S}? For example, traditional arguments against Petschek reconnection fail if the slow mode shocks do not have time to fade. Time-dependent reconnection can generate pulsations in emission which may be used to diagnose the plasma parameters \cite{2023Karampelas,2024Schiavo,2025Schiavo}. What are the observable signatures of oscillatory reconnection in 3D? Can theory predict the periods and damping? How firmly can oscillatory reconnection be connected with quasi-periodic pulsations (QPPs) in flares? 

\subsection*{[Objective 8] Reconnection-Mediated Turbulence and Instabilities}
Turbulent energy cascades are affected by reconnection \citep{2007Loureiro}, which could impact the partition of energy associated with turbulent dissipation. Equally, MMS observations of turbulence in Earth's magnetosheath have revealed novel reconnection dynamics \citep{2018Phan,2022Stawarz}.
What are the properties of reconnection-mediated turbulence and where does it occur? How much energy is dissipated through reconnection in turbulent plasmas and where does it end up? 
A related topic is how reconnection affects the nonlinear development of common instabilities, e.g. by allowing progression of instabilities that would otherwise be stabilised by magnetic tension. How do these govern transport between the solar wind and Earth's magnetosphere, or between open and closed magnetic fields in the solar corona?

\subsection*{[Objective 9] Role in Solar Wind Acceleration and Coronal Heating}
The slow solar wind is believed to originate from interchange reconnection between open and closed fields,  which has been associated with the global \lq{S-web\rq{} of quasi-separatrix layers \citep{2011Antiochos,2023Baker,2025Wilkins} (see white paper by Green). 
At granular scales, SUNRISE has discovered that photospheric magnetic flux is an order of magnitude greater than realised before \citep{2017Smitha}, giving impetus to proposals that reconnection driven by flux emergence, cancellation and braiding contributes to coronal heating and solar wind acceleration \citep{2018Priest,2024Priest,2024Pontin}. Jets of multiple scales also launch waves, flows and switchbacks into the solar wind \citep{2017Wyper,2022Wyper}. Models and detailed observations of these processes are thus a key priority.


\section{UK Leadership, Capability and Partnerships}\label{sec:leadership}

The UK is an international centre of excellence for reconnection that developed many of the fundamental theoretical concepts, e.g. \citep{1958Sweet,1961Dungey}, coined the name \lq{magnetic reconnection}\rq{}, and has a very strong track record in fundamental theory, applications, simulations, observations and instrumentation. The UK contains many of the leading figures in reconnection, who are extremely willing to share knowledge with early career researchers, supporting the future health of the field. The community is exceptionally connected and collaborative, facilitating knowledge exchange between space, solar, astrophysics and fusion research. 
Compared to other countries, the UK has strong competitive advantages in theory and modelling, and in linking theoretical understanding to observations. There is, however, a danger of this advantage slipping without efforts to improve UK funding for theory and mathematical modelling, and to keep UK supercomputers among the top in the world.

The UK has excellent international partnerships on reconnection, including with Europe, the USA, Japan and India. The priorities of this white paper (WP) are closely aligned with  ESA Voyage 2050 \citep{ESAVoyage} themes for medium class missions (see their 3.1.1 to 3.1.4). Reconnection is also centre-stage in all three Science Themes of the US Heliophysics Decadal Survey \citep{DecadalReport}, especially their Guiding Questions
\lq{How Do Fundamental Processes Create and Dissipate Explosive Phenomena Across the Heliosphere?}\rq{} and \lq{How Do Fundamental Processes Govern Coupling Across Spatial Scales?}\rq{}


\section{Priorities: Missions and Hardware}\label{sec:missions}

\paragraph{Plasma Observatory (ESA M7) and Helioswarm}
We strongly back Plasma Observatory (WP by Forsyth) for selection as ESA's M7 mission. A reconnection-focused mission to simultaneously study the ion and fluid scales is the compelling next step after MMS and essential for [Objective 1]. The UK's experience with Cluster, Solar Orbiter and MMS instrumentation and data analysis gives it an excellent leadership position for Plasma Observatory. We also advocate for UK involvement in Helioswarm, which also probes multiple scales but focused on the solar wind turbulence rather than a range of magnetospheric environments [Objective 8]. Since Mars observations are desirable for investigating reconnection in different environments [Objective 4], M-Matisse is our second choice for M7.

\paragraph{SOLAR-C EUVST} (WP by Matthews) is a cornerstone mission for all our science objectives. By producing spectroscopic observations with unprecedented wavelength coverage, diagnostic capability and high resolution, it will probe plasma conditions and structures inside the reconnection region, clarify the roles of shocks and plasmoids, probe energy partitioning by observing the chromospheric response at very high cadence, and characterise  reconnection in partially-ionised plasma. While the mission is already in development, it is important to clearly state that EUVST is of exceptional scientific value and the UK should do its utmost to ensure it comes to full fruition. The UK could also achieve a much greater and more visible leadership role in EUVST across all science areas, including reconnection, if UKSA were to make a small investment in supporting future operations.

\paragraph{SMILE and MUSE Operations and Science}
SMILE (ESA 2026) and MUSE (NASA 2027) will each prove innovative technologies and significantly advance reconnection science. SMILE SXI (Leicester) will image large areas of the Earth’s magnetosheath for the first time, estimating reconnection rates and its spatial extent, and later connect this global perspective with in situ measurements. MUSE will create solar spectrographic maps 35 times faster than predecessors, constraining turbulence, ion heating and the energy inputs to the lower atmosphere during reconnection. While MUSE and SMILE have already been adopted, we wish to explicitly record our full backing and our collective view that high priority should be given to finding additional money to fund UK operations and science.

\paragraph{SPARK (ESA M8)}
A solar flare mission exploring energy release, emission line profiles, and nonthermal particle signatures is strongly aligned with [Objectives 2-7]. The UK-led SPARK (WP by Reid), in Step-2 for ESA's M8 selection, has the backing of the reconnection community. SPARK will build on the advances of EUVST and MUSE, developing new instruments such as an EUV integral field spectrograph that will map line widths and shapes without rastering. SPARK will also play a critical role in progressing our understanding of particle acceleration, determining where and how ions and electrons are accelerated, and where and how they lose their energy.

\paragraph{X-ray Spectroscopy}
A crucial capability not addressed by existing mission concepts is spectroscopic measurements of the hottest material in solar reconnection, where electron temperatures reach about 40~MK. There is also an urgent need for diagnostics that can disambiguate between hot ions and unresolved motions, which could be solved by spatially-resolved spectra for multiple ions with significantly different masses. These requirements point to an X-ray spectrograph that captures hydrogen-like and helium-like lines of ions such as Fe, Ca and S, building on the heritage of Yohkoh's Bragg Crystal Spectrometer \citep{1991Culhane} but with spatial resolution. Options are discussed in the WP by Del Zanna.

\paragraph{UK CubeSats and Sounding Rockets}
Reconnection science goals could be achieved sooner given a faster innovation environment for instruments/missions. In the USA, rapid technological and scientific progress has been made using sounding rockets and CubeSats to fly prototype instruments including X-ray imagers (e.g. FOXSI, WP by Ryan) and spectrographs (e.g. MaGIXS, X-ray WP by Del Zanna). We would be delighted if a similar model were to be successfully rolled out in the UK.

\paragraph{Models in Missions}
As noted in [Pillar 3], models are hugely valuable to hardware development and are a core capability for data interpretation. The approach of funding models within missions (as a relatively cheap quasi-instrument) is an excellent model for future UK-involved missions that amplifies scientific success for a small cost. Critical model requirements are expanded in Section \ref{sec:model-priorities}.

\section{Priorities: Theory, Simulations and Infrastructure}\label{sec:model-priorities}

\paragraph{High-performance Computing (HPC)}
World-leading HPC is essential national infrastructure. For reconnection, scale separation is key to every science objective in Section  \ref{sec:obj}, which makes computing power and advanced codes essential (with features like adaptive mesh refinement or fluid kinetic approaches). As a rule of thumb, doubling 3D resolution  requires a 16$\times$ increase in computing, while doubling system length to increase scale separation requires an 8$\times$ increase. As a benchmark on the \lq{buy in}\rq{} for globally-leading studies, a US simulation of turbulence-mediated reconnection submitted for publication in 2021 had 10000$\times$10000$\times$5000 gridpoints and required 200 million cpu hours \citep{2022Dong}.

\paragraph{Advanced Fluid Modelling}
The UK has world-class expertise in MHD simulations and modelling, including fundamental numerical experiments and solar and magnetospheric modelling. We welcome STFC's investment in the Solar Atmospheric Modelling Suite (SAMS, WP by Hillier), which will catch up UK solar modelling capability with international competitors. 3D modelling of reconnection improves codes, trains users and exploits computing infrastructure. Forward modelling should also be supported as a key capability that underpins comparison between simulations and observations.

\paragraph{Kinetic Modelling}
The UK has a track record of kinetic plasma modelling that includes 2D and 3D PIC simulations, e.g. EPOCH++ code and \citep{2017Franci,2018Franci,2021AgudeloRueda,2022AgudeloRueda,2022Franci_ApJ}. It is important to develop more capacity in kinetic plasma modelling, including large scale 3D models and hybrid (fluid kinetic) frameworks.

\paragraph{Funding Model for Theory}
There are concerns that STFC's Small Awards model is not adequately funding UK theory research. While theory advances represent some of the highest scientific returns, assessors may perceive theory research as riskier than numerics or observations, since it explicitly requires novel ideas during the grant period. Additionally, theory advances may be best achieved by providing established researchers with significant time to think deeply, which does not readily fit a 3-year 100\% RIA + 20\% PL model. Suggestions for improvements include modifying review criteria to better encourage bearing of risk for scientific reward, introducing a stream where theory proposals are assessed by theorists, and allowing higher investigator FTE when RIA time is not requested.

\newpage

\section*{List of Authors and Signatories (Alphabetical)}\vspace{6pt}

\paragraph{Contributions} Co-authors are denoted with asterisks. Authors with a single asterisk * contributed to writing the manuscript through the online survey or email submissions. Authors with two asterisks ** contributed through the survey/email and additionally through editing the manuscript, online discussions etc.\vspace{11pt}

\noindent Keshav Aggarwal \hfill Indian Institute of Technology Indore, India \\
Jeffersson Andres Agudelo Rueda* \hfill Northumbria University, UK \\ 
Oliver Allanson \hfill University of Birmingham, UK \\
Deborah Baker \hfill UCL-MSSL, UK \\ 
William Bate* \hfill University of Hertfordshire, UK \\ 
Ankush Bhaskar \hfill Space Physics Laboratory, VSSC/ISRO, India \\
Shibotosh Biswas \hfill Space Physics Laboratory, VSSC/ISRO, India \\
Yulia Bogdanova \hfill RAL Space, STFC, UKRI, UK \\
Gert Botha \hfill Northumbria University, UK \\ 
Philippa Browning** \hfill University of Manchester, UK \\
Jennifer Carter** \hfill University of Leicester, UK \\
Abhirup Datta \hfill Indian Institute of Technology Indore, India \\
Simon Daley-Yates \hfill University of St Andrews, UK \\
Jordi De Jonghe \hfill KU Leuven, Belgium \\
Giulio Del Zanna* \hfill University of Cambridge, University of Leicester, UK \\
Malcolm Druett (UKSP Chair) \hfill Sheffield University, UK \\
Jonathan Eastwood* \hfill Imperial College London, UK \\
Viktor Fedun \hfill Sheffield University, UK \\
Lyndsay Fletcher \hfill University of Glasgow, UK \\
Luca Franci** \hfill Northumbria University, UK \\
Lucie Green* \hfill UCL-MSSL, UK \\
Iain Hannah \hfill University of Glasgow, UK \\
Laura A. Hayes \hfill Dublin Institute for Advanced Studies, Ireland \\
Heli Hietala** \hfill Queen Mary University of London, UK \\ 
Andrew Hillier** \hfill University of Exeter, UK \\ 
Gunnar Hornig** \hfill University of Dundee, UK \\ 
Konstantinos Karampelas \hfill KU Leuven, Belgium \\
Eduard Kontar \hfill University of Glasgow, UK \\
Marianna Korsos \hfill Sheffield University, UK \\
Anshu Kumari* \hfill Physical Research Laboratory, India \\
Mike Lockwood* (RAS President) \hfill University of Reading, UK \\ 
Duncan Mackay \hfill University of St Andrews, UK \\ 
David MacTaggart** \hfill University of Glasgow, UK \\ 
Sarah Matthews** \hfill UCL-MSSL, UK \\ 
James McKevitt** \hfill UCL-MSSL, UK \\ 
James McLaughlin** \hfill Northumbria University, UK \\ 
Karen Meyer \hfill University of Dundee, UK \\
Teodora Mihailescu \hfill INAF Osservatorio Astronomico di Roma, Italy \\
Qihui Ming \hfill University of St Andrews, UK \\ 
Sargam Mulay \hfill University of Glasgow, UK \\
Thomas Neukirch* \hfill University of St Andrews, UK \\ 
Divya Paliwal \hfill Physical Research Laboratory, India \\ 
Emanuele Papini \hfill INAF Istituto di Astrofisica e Planetologia Spaziali, Italy \\
Clare Parnell \hfill University of St Andrews, UK \\ 
David Pontin* \hfill University of Newcastle, Australia \\ 
Eric Priest** \hfill University of St Andrews, UK \\ 
Hamish Reid \hfill UCL-MSSL, UK \\ 
Jack Reid** \hfill University of St Andrews, UK \\ 
Alexander Russell** (Lead Author and Point of Contact) \hfill University of St Andrews, UK \\ 
Daniel Ryan \hfill UCL-MSSL, UK \\ 
Luiz Schiavo \hfill Northumbria University, UK \\ 
Andy Smith* \hfill Northumbria University, UK \\ 
Ryan Smith \hfill Northumbria University, UK \\ 
Ben Snow** \hfill University of Exeter, UK \\ 
Julia Stawarz** \hfill Northumbria University, UK \\ 
Marzena Trela-Brownlie \hfill University of Southampton, UK \\
Maria-Theresia Walach* \hfill Lancaster University, UK \\
Cara Waters \hfill Queen Mary University of London, UK \\
Antonia Wilmot-Smith \hfill University of St Andrews, UK \\ 
Peter Wyper* \hfill Durham University, UK \\ 
Anthony Yeates** \hfill Durham University, UK \\ 
Natalia Zambrana Prado \hfill UCL-MSSL, UK \\ 
Jinge Zhang \hfill Paris Observatory - PSL, France \\
Valentina Zharkova \hfill Northumbria University, UK

\newpage

\bibliography{main} 

@ARTICLE{2014Amari,
       author = {{Amari}, Tahar and {Canou}, Aur{\'e}lien and {Aly}, Jean-Jacques},
        title = "{Characterizing and predicting the magnetic environment leading to solar eruptions}",
      journal = {\nat},
         year = 2014,
        month = oct,
       volume = {514},
       number = {7523},
        pages = {465-469},
          doi = {10.1038/nature13815},
       adsurl = {https://ui.adsabs.harvard.edu/abs/2014Natur.514..465A}
}

@ARTICLE{2021Antolin,
       author = {{Antolin}, Patrick and {Pagano}, Paolo and {Testa}, Paola and {Petralia}, Antonino and {Reale}, Fabio},
        title = "{Reconnection nanojets in the solar corona}",
      journal = {Nature Astronomy},
         year = 2021,
        month = jan,
       volume = {5},
        pages = {54-62},
          doi = {10.1038/s41550-020-1199-8},
       adsurl = {https://ui.adsabs.harvard.edu/abs/2021NatAs...5...54A}
}

@ARTICLE{2021Arnold,
       author = {{Arnold}, H. and {Drake}, J.~F. and {Swisdak}, M. and {Guo}, F. and {Dahlin}, J.~T. and {Chen}, B. and {Fleishman}, G. and {Glesener}, L. and {Kontar}, E. and {Phan}, T. and {Shen}, C.},
        title = "{Electron Acceleration during Macroscale Magnetic Reconnection}",
      journal = {\prl},
         year = 2021,
        month = apr,
       volume = {126},
       number = {13},
          eid = {135101},
        pages = {135101},
          doi = {10.1103/PhysRevLett.126.135101},
archivePrefix = {arXiv},
       eprint = {2011.01147},
 primaryClass = {physics.plasm-ph},
       adsurl = {https://ui.adsabs.harvard.edu/abs/2021PhRvL.126m5101A}
}

@ARTICLE{2022Beg,
       author = {{Beg}, Raheem and {Russell}, Alexander J.~B. and {Hornig}, Gunnar},
        title = "{Evolution, Structure, and Topology of Self-generated Turbulent Reconnection Layers}",
      journal = {\apj},
         year = 2022,
        month = nov,
       volume = {940},
       number = {1},
          eid = {94},
        pages = {94},
          doi = {10.3847/1538-4357/ac8eb6},
archivePrefix = {arXiv},
       eprint = {2209.04492},
 primaryClass = {astro-ph.SR},
       adsurl = {https://ui.adsabs.harvard.edu/abs/2022ApJ...940...94B}
}

@ARTICLE{2017Beresnyak,
       author = {{Beresnyak}, Andrey},
        title = "{Three-dimensional Spontaneous Magnetic Reconnection}",
      journal = {\apj},
         year = 2017,
        month = jan,
       volume = {834},
       number = {1},
          eid = {47},
        pages = {47},
          doi = {10.3847/1538-4357/834/1/47},
archivePrefix = {arXiv},
       eprint = {1301.7424},
 primaryClass = {astro-ph.SR},
       adsurl = {https://ui.adsabs.harvard.edu/abs/2017ApJ...834...47B}
}

@ARTICLE{2016Burch,
       author = {{Burch}, J.~L. and {Torbert}, R.~B. and {Phan}, T.~D. and {Chen}, L.-J. and {Moore}, T.~E. and {Ergun}, R.~E. and {Eastwood}, J.~P. and {Gershman}, D.~J. and {Cassak}, P.~A. and {Argall}, M.~R. and {Wang}, S. and {Hesse}, M. and {Pollock}, C.~J. and {Giles}, B.~L. and {Nakamura}, R. and {Mauk}, B.~H. and {Fuselier}, S.~A. and {Russell}, C.~T. and {Strangeway}, R.~J. and {Drake}, J.~F. and {Shay}, M.~A. and {Khotyaintsev}, Yu. V. and {Lindqvist}, P.-A. and {Marklund}, G. and {Wilder}, F.~D. and {Young}, D.~T. and {Torkar}, K. and {Goldstein}, J. and {Dorelli}, J.~C. and {Avanov}, L.~A. and {Oka}, M. and {Baker}, D.~N. and {Jaynes}, A.~N. and {Goodrich}, K.~A. and {Cohen}, I.~J. and {Turner}, D.~L. and {Fennell}, J.~F. and {Blake}, J.~B. and {Clemmons}, J. and {Goldman}, M. and {Newman}, D. and {Petrinec}, S.~M. and {Trattner}, K.~J. and {Lavraud}, B. and {Reiff}, P.~H. and {Baumjohann}, W. and {Magnes}, W. and {Steller}, M. and {Lewis}, W. and {Saito}, Y. and {Coffey}, V. and {Chandler}, M.},
        title = "{Electron-scale measurements of magnetic reconnection in space}",
      journal = {Science},
         year = 2016,
        month = jun,
       volume = {352},
          eid = {aaf2939},
        pages = {aaf2939},
          doi = {10.1126/science.aaf2939},
       adsurl = {https://ui.adsabs.harvard.edu/abs/2016Sci...352.2939B}
}

@ARTICLE{2015Chen,
       author = {{Chen}, Bin and {Bastian}, Timothy S. and {Shen}, Chengcai and {Gary}, Dale E. and {Krucker}, S{\"a}m and {Glesener}, Lindsay},
        title = "{Particle acceleration by a solar flare termination shock}",
      journal = {Science},
         year = 2015,
        month = dec,
       volume = {350},
       number = {6265},
        pages = {1238-1242},
          doi = {10.1126/science.aac8467},
archivePrefix = {arXiv},
       eprint = {1512.02237},
 primaryClass = {astro-ph.SR},
       adsurl = {https://ui.adsabs.harvard.edu/abs/2015Sci...350.1238C}
}

@ARTICLE{2018Cheng,
       author = {{Cheng}, X. and {Li}, Y. and {Wan}, L.~F. and {Ding}, M.~D. and {Chen}, P.~F. and {Zhang}, J. and {Liu}, J.~J.},
        title = "{Observations of Turbulent Magnetic Reconnection within a Solar Current Sheet}",
      journal = {\apj},
         year = 2018,
        month = oct,
       volume = {866},
       number = {1},
          eid = {64},
        pages = {64},
          doi = {10.3847/1538-4357/aadd16},
archivePrefix = {arXiv},
       eprint = {1808.06071},
 primaryClass = {astro-ph.SR},
       adsurl = {https://ui.adsabs.harvard.edu/abs/2018ApJ...866...64C}
}

@ARTICLE{2023Cheng,
       author = {{Cheng}, X. and {Priest}, E.~R. and {Li}, H.~T. and {Chen}, J. and {Aulanier}, G. and {Chitta}, L.~P. and {Wang}, Y.~L. and {Peter}, H. and {Zhu}, X.~S. and {Xing}, C. and {Ding}, M.~D. and {Solanki}, S.~K. and {Berghmans}, D. and {Teriaca}, L. and {Aznar Cuadrado}, R. and {Zhukov}, A.~N. and {Guo}, Y. and {Long}, D. and {Harra}, L. and {Smith}, P.~J. and {Rodriguez}, L. and {Verbeeck}, C. and {Barczynski}, K. and {Parenti}, S.},
        title = "{Ultra-high-resolution observations of persistent null-point reconnection in the solar corona}",
      journal = {Nature Communications},
         year = 2023,
        month = apr,
       volume = {14},
          eid = {2107},
        pages = {2107},
          doi = {10.1038/s41467-023-37888-w},
archivePrefix = {arXiv},
       eprint = {2304.08725},
 primaryClass = {astro-ph.SR},
       adsurl = {https://ui.adsabs.harvard.edu/abs/2023NatCo..14.2107C}
}

@ARTICLE{2022Chitta,
       author = {{Chitta}, L.~P. and {Peter}, H. and {Parenti}, S. and {Berghmans}, D. and {Auch{\`e}re}, F. and {Solanki}, S.~K. and {Aznar Cuadrado}, R. and {Sch{\"u}hle}, U. and {Teriaca}, L. and {Mandal}, S. and {Barczynski}, K. and {Buchlin}, {\'E}. and {Harra}, L. and {Kraaikamp}, E. and {Long}, D.~M. and {Rodriguez}, L. and {Schwanitz}, C. and {Smith}, P.~J. and {Verbeeck}, C. and {Zhukov}, A.~N. and {Liu}, W. and {Cheung}, M.~C.~M.},
        title = "{Solar coronal heating from small-scale magnetic braids}",
      journal = {\aap},
         year = 2022,
        month = nov,
       volume = {667},
          eid = {A166},
        pages = {A166},
          doi = {10.1051/0004-6361/202244170},
archivePrefix = {arXiv},
       eprint = {2209.12203},
 primaryClass = {astro-ph.SR},
       adsurl = {https://ui.adsabs.harvard.edu/abs/2022A&A...667A.166C}
}

@ARTICLE{1991Culhane,
       author = {{Culhane}, J.~L. and {Hiei}, E. and {Doschek}, G.~A. and {Cruise}, A.~M. and {Ogawara}, Y. and {Uchida}, Y. and {Bentley}, R.~D. and {Brown}, C.~M. and {Lang}, J. and {Watanabe}, T. and {Bowles}, J.~A. and {Deslattes}, R.~D. and {Feldman}, U. and {Fludra}, A. and {Guttridge}, P. and {Henins}, A. and {Lapington}, J. and {Magraw}, J. and {Mariska}, J.~T. and {Payne}, J. and {Phillips}, K.~J.~H. and {Sheather}, P. and {Slater}, K. and {Tanaka}, K. and {Towndrow}, E. and {Trow}, M.~W. and {Yamaguchi}, A.},
        title = "{The Bragg Crystal Spectrometer for SOLAR-A}",
      journal = {\solphys},
         year = 1991,
        month = nov,
       volume = {136},
       number = {1},
        pages = {89-104},
          doi = {10.1007/BF00151696},
       adsurl = {https://ui.adsabs.harvard.edu/abs/1991SoPh..136...89C}
}

@ARTICLE{2025Desai,
       author = {{Desai}, M.~I. and {Drake}, J.~F. and {Phan}, T. and {Yin}, Z. and {Swisdak}, M. and {McComas}, D.~J. and {Bale}, S.~D. and {Rahmati}, A. and {Larson}, D. and {Matthaeus}, W.~H. and {Dayeh}, M.~A. and {Starkey}, M.~J. and {Raouafi}, N.~E. and {Mitchell}, D.~G. and {Cohen}, C.~M.~S. and {Szalay}, J.~R. and {Giacalone}, J. and {Hill}, M.~E. and {Christian}, E.~R. and {Schwadron}, N.~A. and {McNutt}, R.~L. and {Malandraki}, O. and {Whittlesey}, P. and {Livi}, R. and {Kasper}, J.~C.},
        title = "{Magnetic Reconnection{\textendash}driven Energization of Protons up to {\ensuremath{\sim}}400 keV at the Near-Sun Heliospheric Current Sheet}",
      journal = {\apjl},
         year = 2025,
        month = jun,
       volume = {985},
       number = {2},
          eid = {L38},
        pages = {L38},
          doi = {10.3847/2041-8213/ada697},
archivePrefix = {arXiv},
       eprint = {2410.16539},
 primaryClass = {astro-ph.SR},
       adsurl = {https://ui.adsabs.harvard.edu/abs/2025ApJ...985L..38D}
}

@ARTICLE{2022Dong,
       author = {{Dong}, Chuanfei and {Wang}, Liang and {Huang}, Yi-Min and {Comisso}, Luca and {Sandstrom}, Timothy A. and {Bhattacharjee}, Amitava},
        title = "{Reconnection-driven energy cascade in magnetohydrodynamic turbulence}",
      journal = {Science Advances},
         year = 2022,
        month = dec,
       volume = {8},
       number = {49},
          eid = {eabn7627},
        pages = {eabn7627},
          doi = {10.1126/sciadv.abn7627},
archivePrefix = {arXiv},
       eprint = {2210.10736},
 primaryClass = {astro-ph.SR},
       adsurl = {https://ui.adsabs.harvard.edu/abs/2022SciA....8N7627D}
}

@ARTICLE{2011Daughton,
       author = {{Daughton}, W. and {Roytershteyn}, V. and {Karimabadi}, H. and {Yin}, L. and {Albright}, B.~J. and {Bergen}, B. and {Bowers}, K.~J.},
        title = "{Role of electron physics in the development of turbulent magnetic reconnection in collisionless plasmas}",
      journal = {Nature Physics},
         year = 2011,
        month = jul,
       volume = {7},
       number = {7},
        pages = {539-542},
          doi = {10.1038/nphys1965},
       adsurl = {https://ui.adsabs.harvard.edu/abs/2011NatPh...7..539D}
}

@ARTICLE{2014Daughton,
       author = {{Daughton}, W. and {Nakamura}, T.~K.~M. and {Karimabadi}, H. and {Roytershteyn}, V. and {Loring}, B.},
        title = "{Computing the reconnection rate in turbulent kinetic layers by using electron mixing to identify topology}",
      journal = {Physics of Plasmas},
         year = 2014,
        month = may,
       volume = {21},
       number = {5},
          eid = {052307},
        pages = {052307},
          doi = {10.1063/1.4875730},
       adsurl = {https://ui.adsabs.harvard.edu/abs/2014PhPl...21e2307D}
}

@ARTICLE{2009Drake,
       author = {{Drake}, J.~F. and {Swisdak}, M. and {Phan}, T.~D. and {Cassak}, P.~A. and {Shay}, M.~A. and {Lepri}, S.~T. and {Lin}, R.~P. and {Quataert}, E. and {Zurbuchen}, T.~H.},
        title = "{Ion heating resulting from pickup in magnetic reconnection exhausts}",
      journal = {Journal of Geophysical Research (Space Physics)},
         year = 2009,
        month = may,
       volume = {114},
       number = {A5},
          eid = {A05111},
        pages = {A05111},
          doi = {10.1029/2008JA013701},
       adsurl = {https://ui.adsabs.harvard.edu/abs/2009JGRA..114.5111D}
}

@ARTICLE{1961Dungey,
       author = {{Dungey}, J.~W.},
        title = "{Interplanetary Magnetic Field and the Auroral Zones}",
      journal = {\prl},
         year = 1961,
        month = jan,
       volume = {6},
       number = {2},
        pages = {47-48},
          doi = {10.1103/PhysRevLett.6.47},
       adsurl = {https://ui.adsabs.harvard.edu/abs/1961PhRvL...6...47D}
}

@ARTICLE{2013Eastwood,
       author = {{Eastwood}, J.~P. and {Phan}, T.~D. and {Drake}, J.~F. and {Shay}, M.~A. and {Borg}, A.~L. and {Lavraud}, B. and {Taylor}, M.~G.~G.~T.},
        title = "{Energy Partition in Magnetic Reconnection in Earth's Magnetotail}",
      journal = {\prl},
         year = 2013,
        month = may,
       volume = {110},
       number = {22},
          eid = {225001},
        pages = {225001},
          doi = {10.1103/PhysRevLett.110.225001},
       adsurl = {https://ui.adsabs.harvard.edu/abs/2013PhRvL.110v5001E}
}

@ARTICLE{2020Eastwood,
       author = {{Eastwood}, J.~P. and {Goldman}, M.~V. and {Phan}, T.~D. and {Stawarz}, J.~E. and {Cassak}, P.~A. and {Drake}, J.~F. and {Newman}, D. and {Lavraud}, B. and {Shay}, M.~A. and {Ergun}, R.~E. and {Burch}, J.~L. and {Gershman}, D.~J. and {Giles}, B.~L. and {Lindqvist}, P.~A. and {Torbert}, R.~B. and {Strangeway}, R.~J. and {Russell}, C.~T.},
        title = "{Energy Flux Densities near the Electron Dissipation Region in Asymmetric Magnetopause Reconnection}",
      journal = {\prl},
         year = 2020,
        month = dec,
       volume = {125},
       number = {26},
          eid = {265102},
        pages = {265102},
          doi = {10.1103/PhysRevLett.125.265102},
       adsurl = {https://ui.adsabs.harvard.edu/abs/2020PhRvL.125z5102E}
}

@ARTICLE{2024Fargette,
       author = {{Fargette}, Na{\"\i}s. and {Eastwood}, Jonathan P. and {Waters}, Cara L. and {{\O}ieroset}, Marit and {Phan}, Tai D. and {Newman}, David L. and {Stawarz}, J.~E. and {Goldman}, Martin V. and {Lapenta}, Giovanni},
        title = "{Statistical Study of Energy Transport and Conversion in Electron Diffusion Regions at Earth's Dayside Magnetopause}",
      journal = {Journal of Geophysical Research (Space Physics)},
         year = 2024,
        month = oct,
       volume = {129},
       number = {10},
          eid = {e2024JA032897},
        pages = {e2024JA032897},
          doi = {10.1029/2024JA032897},
       adsurl = {https://ui.adsabs.harvard.edu/abs/2024JGRA..12932897F}
}

@ARTICLE{2022Fleishman,
       author = {{Fleishman}, Gregory D. and {Nita}, Gelu M. and {Chen}, Bin and {Yu}, Sijie and {Gary}, Dale E.},
        title = "{Solar flare accelerates nearly all electrons in a large coronal volume}",
      journal = {\nat},
         year = 2022,
        month = jun,
       volume = {606},
       number = {7915},
        pages = {674-677},
          doi = {10.1038/s41586-022-04728-8},
       adsurl = {https://ui.adsabs.harvard.edu/abs/2022Natur.606..674F}
}

@ARTICLE{2011Fletcher,
       author = {{Fletcher}, L. and {Dennis}, B.~R. and {Hudson}, H.~S. and {Krucker}, S. and {Phillips}, K. and {Veronig}, A. and {Battaglia}, M. and {Bone}, L. and {Caspi}, A. and {Chen}, Q. and {Gallagher}, P. and {Grigis}, P.~T. and {Ji}, H. and {Liu}, W. and {Milligan}, R.~O. and {Temmer}, M.},
        title = "{An Observational Overview of Solar Flares}",
      journal = {\ssr},
         year = 2011,
        month = sep,
       volume = {159},
       number = {1-4},
        pages = {19-106},
          doi = {10.1007/s11214-010-9701-8},
archivePrefix = {arXiv},
       eprint = {1109.5932},
 primaryClass = {astro-ph.SR},
       adsurl = {https://ui.adsabs.harvard.edu/abs/2011SSRv..159...19F}
}

@ARTICLE{2021French,
       author = {{French}, Ryan J. and {Matthews}, Sarah A. and {Jonathan Rae}, I. and {Smith}, Andrew W.},
        title = "{Probing Current Sheet Instabilities from Flare Ribbon Dynamics}",
      journal = {\apj},
         year = 2021,
        month = dec,
       volume = {922},
       number = {2},
          eid = {117},
        pages = {117},
          doi = {10.3847/1538-4357/ac256f},
archivePrefix = {arXiv},
       eprint = {2109.03753},
 primaryClass = {astro-ph.SR},
       adsurl = {https://ui.adsabs.harvard.edu/abs/2021ApJ...922..117F}
}

@ARTICLE{2024French,
       author = {{French}, Ryan J. and {Yu}, Sijie and {Chen}, Bin and {Shen}, Chengcai and {Matthews}, Sarah A.},
        title = "{Doppler signature of a possible termination shock in an off-limb solar flare}",
      journal = {\mnras},
         year = 2024,
        month = mar,
       volume = {528},
       number = {4},
        pages = {6836-6844},
          doi = {10.1093/mnras/stae430},
archivePrefix = {arXiv},
       eprint = {2402.04445},
 primaryClass = {astro-ph.SR},
       adsurl = {https://ui.adsabs.harvard.edu/abs/2024MNRAS.528.6836F}
}

@ARTICLE{2021Froment,
       author = {{Froment}, C. and {Krasnoselskikh}, V. and {Dudok de Wit}, T. and {Agapitov}, O. and {Fargette}, N. and {Lavraud}, B. and {Larosa}, A. and {Kretzschmar}, M. and {Jagarlamudi}, V.~K. and {Velli}, M. and {Malaspina}, D. and {Whittlesey}, P.~L. and {Bale}, S.~D. and {Case}, A.~W. and {Goetz}, K. and {Kasper}, J.~C. and {Korreck}, K.~E. and {Larson}, D.~E. and {MacDowall}, R.~J. and {Mozer}, F.~S. and {Pulupa}, M. and {Revillet}, C. and {Stevens}, M.~L.},
        title = "{Direct evidence for magnetic reconnection at the boundaries of magnetic switchbacks with Parker Solar Probe}",
      journal = {\aap},
         year = 2021,
        month = jun,
       volume = {650},
          eid = {A5},
        pages = {A5},
          doi = {10.1051/0004-6361/202039806},
archivePrefix = {arXiv},
       eprint = {2101.06279},
 primaryClass = {astro-ph.SR},
       adsurl = {https://ui.adsabs.harvard.edu/abs/2021A&A...650A...5F}
}

@ARTICLE{2015Graham,
       author = {{Graham}, D.~R. and {Cauzzi}, G.},
        title = "{Temporal Evolution of Multiple Evaporating Ribbon Sources in a Solar Flare}",
      journal = {\apjl},
         year = 2015,
        month = jul,
       volume = {807},
       number = {2},
          eid = {L22},
        pages = {L22},
          doi = {10.1088/2041-8205/807/2/L22},
archivePrefix = {arXiv},
       eprint = {1506.03465},
 primaryClass = {astro-ph.SR},
       adsurl = {https://ui.adsabs.harvard.edu/abs/2015ApJ...807L..22G}
}

@ARTICLE{2025Haahr,
       author = {{Haahr}, M. and {Gudiksen}, B.~V. and {Nordlund}, {\r{A}}.},
        title = "{Coupling particle-in-cell and magnetohydrodynamics methods for realistic solar flare models}",
      journal = {\aap},
         year = 2025,
        month = apr,
       volume = {696},
          eid = {A191},
        pages = {A191},
          doi = {10.1051/0004-6361/202452117},
archivePrefix = {arXiv},
       eprint = {2409.02493},
 primaryClass = {astro-ph.SR},
       adsurl = {https://ui.adsabs.harvard.edu/abs/2025A&A...696A.191H}
}

@ARTICLE{2020HesseCassak,
       author = {{Hesse}, M. and {Cassak}, P.~A.},
        title = "{Magnetic Reconnection in the Space Sciences: Past, Present, and Future}",
      journal = {Journal of Geophysical Research (Space Physics)},
         year = 2020,
        month = feb,
       volume = {125},
       number = {2},
          eid = {e25935},
        pages = {e25935},
          doi = {10.1029/2018JA025935},
       adsurl = {https://ui.adsabs.harvard.edu/abs/2020JGRA..12525935H}
}

@ARTICLE{2018Hoshino,
       author = {{Hoshino}, Masahiro},
        title = "{Energy Partition between Ion and Electron of Collisionless Magnetic Reconnection}",
      journal = {\apjl},
         year = 2018,
        month = dec,
       volume = {868},
       number = {2},
          eid = {L18},
        pages = {L18},
          doi = {10.3847/2041-8213/aaef3a},
archivePrefix = {arXiv},
       eprint = {1811.03226},
 primaryClass = {astro-ph.HE},
       adsurl = {https://ui.adsabs.harvard.edu/abs/2018ApJ...868L..18H}
}

@ARTICLE{2016HuangBhattacharjee,
       author = {{Huang}, Yi-Min and {Bhattacharjee}, A.},
        title = "{Turbulent Magnetohydrodynamic Reconnection Mediated by the Plasmoid Instability}",
      journal = {\apj},
         year = 2016,
        month = feb,
       volume = {818},
       number = {1},
          eid = {20},
        pages = {20},
          doi = {10.3847/0004-637X/818/1/20},
archivePrefix = {arXiv},
       eprint = {1512.01520},
 primaryClass = {physics.plasm-ph},
       adsurl = {https://ui.adsabs.harvard.edu/abs/2016ApJ...818...20H}
}

@ARTICLE{2024Huang,
       author = {{Huang}, Yi-Min and {Bhattacharjee}, Amitava},
        title = "{Three-dimensional plasmoid-mediated reconnection and turbulence in Hall magnetohydrodynamics}",
      journal = {Physics of Plasmas},
         year = 2024,
        month = aug,
       volume = {31},
       number = {8},
          eid = {082119},
        pages = {082119},
          doi = {10.1063/5.0216561},
archivePrefix = {arXiv},
       eprint = {2404.19285},
 primaryClass = {physics.plasm-ph},
       adsurl = {https://ui.adsabs.harvard.edu/abs/2024PhPl...31h2119H}
}

@ARTICLE{2013Imada,
       author = {{Imada}, S. and {Aoki}, K. and {Hara}, H. and {Watanabe}, T. and {Harra}, L.~K. and {Shimizu}, T.},
        title = "{Evidence for Hot Fast Flow above a Solar Flare Arcade}",
      journal = {\apjl},
         year = 2013,
        month = oct,
       volume = {776},
       number = {1},
          eid = {L11},
        pages = {L11},
          doi = {10.1088/2041-8205/776/1/L11},
archivePrefix = {arXiv},
       eprint = {1309.3401},
 primaryClass = {astro-ph.SR},
       adsurl = {https://ui.adsabs.harvard.edu/abs/2013ApJ...776L..11I}
}

@ARTICLE{2011JiDaughton,
       author = {{Ji}, Hantao and {Daughton}, William},
        title = "{Phase diagram for magnetic reconnection in heliophysical, astrophysical, and laboratory plasmas}",
      journal = {Physics of Plasmas},
         year = 2011,
        month = nov,
       volume = {18},
       number = {11},
        pages = {111207-111207},
          doi = {10.1063/1.3647505},
archivePrefix = {arXiv},
       eprint = {1109.0756},
 primaryClass = {astro-ph.IM},
       adsurl = {https://ui.adsabs.harvard.edu/abs/2011PhPl...18k1207J}
}

@ARTICLE{2022Ji,
       author = {{Ji}, Hantao and {Daughton}, William and {Jara-Almonte}, Jonathan and {Le}, Ari and {Stanier}, Adam and {Yoo}, Jongsoo},
        title = "{Magnetic reconnection in the era of exascale computing and multiscale experiments}",
      journal = {Nature Reviews Physics},
         year = 2022,
        month = apr,
       volume = {4},
       number = {4},
        pages = {263-282},
          doi = {10.1038/s42254-021-00419-x},
archivePrefix = {arXiv},
       eprint = {2202.09004},
 primaryClass = {physics.plasm-ph},
       adsurl = {https://ui.adsabs.harvard.edu/abs/2022NatRP...4..263J}
}

@ARTICLE{2023Ji,
       author = {{Ji}, H. and {Yoo}, J. and {Fox}, W. and {Yamada}, M. and {Argall}, M. and {Egedal}, J. and {Liu}, Y.-H. and {Wilder}, R. and {Eriksson}, S. and {Daughton}, W. and {Bergstedt}, K. and {Bose}, S. and {Burch}, J. and {Torbert}, R. and {Ng}, J. and {Chen}, L.-J.},
        title = "{Laboratory Study of Collisionless Magnetic Reconnection}",
      journal = {\ssr},
         year = 2023,
        month = dec,
       volume = {219},
       number = {8},
          eid = {76},
        pages = {76},
          doi = {10.1007/s11214-023-01024-3},
archivePrefix = {arXiv},
       eprint = {2307.07109},
 primaryClass = {physics.plasm-ph},
       adsurl = {https://ui.adsabs.harvard.edu/abs/2023SSRv..219...76J}
}

@INPROCEEDINGS{2023Decadal,
       author = {{Ji}, Hantao and {Karpen}, Judth and {Alt}, Andrew and {Bellan}, Paul and {Begelman}, Mitch and {Beresnyak}, Andrey and {Blackman}, Eric and {Bose}, Sayak and {Brown}, Mike and {Burch}, James and {Carter}, Troy and {Cassak}, Paul and {Chen}, Bin and {Chen}, Li-Jen and {Cheung}, Mark and {Comisso}, Luca and {Dahlin}, Joel and {Daughton}, William and {DeLuca}, Ed and {Dong}, Chuanfei and {Dorfman}, Seth and {Drake}, JIm and {Ebrahimi}, Fatima and {Egedal}, Jan and {Forest}, Cary and {Froula}, Dustin and {Fujimoto}, Keizo and {Gao}, Lan and {Genestreti}, Kevin and {Gibson}, Sarah and {Guo}, Fan and {Hoshino}, Masahiro and {Hu}, Qiang and {Huang}, Yi-Min and {Karimabadi}, Homa and {Kepco}, Larry and {Klimchuk}, Jim and {Kunz}, Matthew and {Kusano}, Kanya and {Lazarian}, Alex and {Lebedev}, Sergey and {Li}, Hui and {Li}, Xiaocan and {Lin}, Yu and {Linton}, Mark and {Liu}, Yi-Hsin and {Loureiro}, Nuno and {Majeski}, Stephen and {Matthaeus}, William and {McLaughlin}, James and {Murphy}, Nick and {Ono}, Yasushi and {Opher}, Merav and {Qiu}, Jiong and {Rempel}, Matthias and {Ren}, Yang and {Rosner}, Robert and {Roytershteyn}, Vadim and {Savcheva}, Antonia and {Schoeffier}, Kevin and {Scime}, Earl and {Shi}, Peiyun and {Sironi}, Lorenzo and {Stanier}, Adam and {TenBarge}, Jason and {Vaivads}, Andris and {Wang}, Haimin and {Yamada}, Masaaki and {Yokoyama}, Takaaki and {Yoo}, Jongsoo and {Zenitani}, Seiji and {Zhang}, Jie and {Zweibel}, Ellen},
        title = "{Major Scientific Challenges and Opportunities in Understanding Magnetic Reconnection and Related Explosive Phenomena in Heliophysics and Beyond}",
    booktitle = {Bulletin of the American Astronomical Society},
         year = 2023,
       volume = {55},
        month = jul,
          eid = {192},
        pages = {192},
          doi = {10.3847/25c2cfeb.e22a8d1f},
       adsurl = {https://ui.adsabs.harvard.edu/abs/2023BAAS...55c.192J}
}

@ARTICLE{2017Kontar,
       author = {{Kontar}, E.~P. and {Perez}, J.~E. and {Harra}, L.~K. and {Kuznetsov}, A.~A. and {Emslie}, A.~G. and {Jeffrey}, N.~L.~S. and {Bian}, N.~H. and {Dennis}, B.~R.},
        title = "{Turbulent Kinetic Energy in the Energy Balance of a Solar Flare}",
      journal = {\prl},
         year = 2017,
        month = apr,
       volume = {118},
       number = {15},
          eid = {155101},
        pages = {155101},
          doi = {10.1103/PhysRevLett.118.155101},
archivePrefix = {arXiv},
       eprint = {1703.02392},
 primaryClass = {astro-ph.SR},
       adsurl = {https://ui.adsabs.harvard.edu/abs/2017PhRvL.118o5101K}
}

@ARTICLE{2010Krucker,
       author = {{Krucker}, S{\"a}m and {Hudson}, H.~S. and {Glesener}, L. and {White}, S.~M. and {Masuda}, S. and {Wuelser}, J.-P. and {Lin}, R.~P.},
        title = "{Measurements of the Coronal Acceleration Region of a Solar Flare}",
      journal = {\apj},
         year = 2010,
        month = may,
       volume = {714},
       number = {2},
        pages = {1108-1119},
          doi = {10.1088/0004-637X/714/2/1108},
       adsurl = {https://ui.adsabs.harvard.edu/abs/2010ApJ...714.1108K}
}

@ARTICLE{2014Krucker,
       author = {{Krucker}, S{\"a}m and {Battaglia}, Marina},
        title = "{Particle Densities within the Acceleration Region of a Solar Flare}",
      journal = {\apj},
         year = 2014,
        month = jan,
       volume = {780},
       number = {1},
          eid = {107},
        pages = {107},
          doi = {10.1088/0004-637X/780/1/107},
       adsurl = {https://ui.adsabs.harvard.edu/abs/2014ApJ...780..107K}
}

@ARTICLE{2024Kumar,
       author = {{Kumar}, Pankaj and {Nakariakov}, Valery M. and {Karpen}, Judith T. and {Cho}, Kyung-Suk},
        title = "{Direct imaging of magnetohydrodynamic wave mode conversion near a 3D null point on the sun}",
      journal = {Nature Communications},
         year = 2024,
        month = mar,
       volume = {15},
          eid = {2667},
        pages = {2667},
          doi = {10.1038/s41467-024-46736-4},
archivePrefix = {arXiv},
       eprint = {2403.02250},
 primaryClass = {astro-ph.SR},
       adsurl = {https://ui.adsabs.harvard.edu/abs/2024NatCo..15.2667K}
}

@ARTICLE{2018Kuramitsu,
       author = {{Kuramitsu}, Y. and {Moritaka}, T. and {Sakawa}, Y. and {Morita}, T. and {Sano}, T. and {Koenig}, M. and {Gregory}, C.~D. and {Woolsey}, N. and {Tomita}, K. and {Takabe}, H. and {Liu}, Y.~L. and {Chen}, S.~H. and {Matsukiyo}, S. and {Hoshino}, M.},
        title = "{Magnetic reconnection driven by electron dynamics}",
      journal = {Nature Communications},
         year = 2018,
        month = nov,
       volume = {9},
          eid = {5109},
        pages = {5109},
          doi = {10.1038/s41467-018-07415-3},
       adsurl = {https://ui.adsabs.harvard.edu/abs/2018NatCo...9.5109K}
}

@ARTICLE{2018Li,
       author = {{Li}, Y. and {Xue}, J.~C. and {Ding}, M.~D. and {Cheng}, X. and {Su}, Y. and {Feng}, L. and {Hong}, J. and {Li}, H. and {Gan}, W.~Q.},
        title = "{Spectroscopic Observations of a Current Sheet in a Solar Flare}",
      journal = {\apjl},
         year = 2018,
        month = jan,
       volume = {853},
       number = {1},
          eid = {L15},
        pages = {L15},
          doi = {10.3847/2041-8213/aaa6c0},
archivePrefix = {arXiv},
       eprint = {1801.03631},
 primaryClass = {astro-ph.SR},
       adsurl = {https://ui.adsabs.harvard.edu/abs/2018ApJ...853L..15L}
}

@ARTICLE{2007Loureiro,
       author = {{Loureiro}, N.~F. and {Schekochihin}, A.~A. and {Cowley}, S.~C.},
        title = "{Instability of current sheets and formation of plasmoid chains}",
      journal = {Physics of Plasmas},
         year = 2007,
        month = oct,
       volume = {14},
       number = {10},
        pages = {100703-100703},
          doi = {10.1063/1.2783986},
archivePrefix = {arXiv},
       eprint = {astro-ph/0703631},
 primaryClass = {astro-ph},
       adsurl = {https://ui.adsabs.harvard.edu/abs/2007PhPl...14j0703L}
}

@ARTICLE{2025MacTaggart,
       author = {{MacTaggart}, David},
        title = "{On Field Line Slippage Rates in the Solar Corona}",
      journal = {\solphys},
         year = 2025,
        month = apr,
       volume = {300},
       number = {4},
          eid = {48},
        pages = {48},
          doi = {10.1007/s11207-025-02462-8},
archivePrefix = {arXiv},
       eprint = {2502.01251},
 primaryClass = {astro-ph.SR},
       adsurl = {https://ui.adsabs.harvard.edu/abs/2025SoPh..300...48M}
}

@unpublished{2026Ming,
  author = {{Ming}, Q. and {Russell}, A.~J.~B.},
  title = {Distinct Anti-parallel and Guide Field Regimes of {MHD} Self-Generated Turbulent Reconnection},
  note = {{Manuscript in preparation}}
}

@ARTICLE{2017Misty,
       author = {{Mistry}, R. and {Eastwood}, J.~P. and {Phan}, T.~D. and {Hietala}, H.},
        title = "{Statistical properties of solar wind reconnection exhausts}",
      journal = {Journal of Geophysical Research (Space Physics)},
         year = 2017,
        month = jun,
       volume = {122},
       number = {6},
        pages = {5895-5909},
          doi = {10.1002/2017JA024032},
       adsurl = {https://ui.adsabs.harvard.edu/abs/2017JGRA..122.5895M}
}

@ARTICLE{2025Nakamura,
       author = {{Nakamura}, R. and {Burch}, J.~L. and {Birn}, J. and {Chen}, L.-J. and {Graham}, D.~B. and {Guo}, F. and {Hwang}, K.-J. and {Ji}, H. and {Khotyaintsev}, Y.~V. and {Liu}, Y.-H. and {Oka}, M. and {Payne}, D. and {Sitnov}, M.~I. and {Swisdak}, M. and {Zenitani}, S. and {Drake}, J.~F. and {Fuselier}, S.~A. and {Genestreti}, K.~J. and {Gershman}, D.~J. and {Hasegawa}, H. and {Hoshino}, M. and {Norgren}, C. and {Shay}, M.~A. and {Shuster}, J.~R. and {Stawarz}, J.~E.},
        title = "{Outstanding Questions and Future Research on Magnetic Reconnection}",
      journal = {\ssr},
         year = 2025,
        month = feb,
       volume = {221},
       number = {1},
          eid = {17},
        pages = {17},
          doi = {10.1007/s11214-025-01143-z},
archivePrefix = {arXiv},
       eprint = {2407.09670},
 primaryClass = {physics.plasm-ph},
       adsurl = {https://ui.adsabs.harvard.edu/abs/2025SSRv..221...17N}
}

@ARTICLE{2015Oishi,
       author = {{Oishi}, Jeffrey S. and {Mac Low}, Mordecai-Mark and {Collins}, David C. and {Tamura}, Moeko},
        title = "{Self-generated Turbulence in Magnetic Reconnection}",
      journal = {\apjl},
         year = 2015,
        month = jun,
       volume = {806},
       number = {1},
          eid = {L12},
        pages = {L12},
          doi = {10.1088/2041-8205/806/1/L12},
archivePrefix = {arXiv},
       eprint = {1505.04653},
 primaryClass = {astro-ph.SR},
       adsurl = {https://ui.adsabs.harvard.edu/abs/2015ApJ...806L..12O}
}

@ARTICLE{2025Oka,
       author = {{Oka}, Mitsuo and {Phan}, Tai D. and {{\O}ieroset}, Marit and {Gershman}, Daniel J. and {Torbert}, Roy B. and {Burch}, James L. and {Angelopoulos}, Vassilis},
        title = "{Scaling of Particle Heating in Shocks and Magnetic Reconnection}",
      journal = {\apj},
         year = 2025,
        month = may,
       volume = {984},
       number = {2},
          eid = {150},
        pages = {150},
          doi = {10.3847/1538-4357/adc5e5},
archivePrefix = {arXiv},
       eprint = {2503.14823},
 primaryClass = {physics.plasm-ph},
       adsurl = {https://ui.adsabs.harvard.edu/abs/2025ApJ...984..150O}
}

@ARTICLE{2018Omodei,
       author = {{Omodei}, Nicola and {Pesce-Rollins}, Melissa and {Longo}, Francesco and {Allafort}, Alice and {Krucker}, S{\"a}m},
        title = "{Fermi-LAT Observations of the 2017 September 10 Solar Flare}",
      journal = {\apjl},
         year = 2018,
        month = sep,
       volume = {865},
       number = {1},
          eid = {L7},
        pages = {L7},
          doi = {10.3847/2041-8213/aae077},
archivePrefix = {arXiv},
       eprint = {1803.07654},
 primaryClass = {astro-ph.HE},
       adsurl = {https://ui.adsabs.harvard.edu/abs/2018ApJ...865L...7O}
}

@ARTICLE{2024Parnell,
       author = {{Parnell}, C.~E.},
        title = "{On the importance of separators as sites of 3D magnetic reconnection}",
      journal = {Physics of Plasmas},
         year = 2024,
        month = aug,
       volume = {31},
       number = {8},
          eid = {082112},
        pages = {082112},
          doi = {10.1063/5.0189787},
archivePrefix = {arXiv},
       eprint = {2403.04076},
 primaryClass = {astro-ph.SR},
       adsurl = {https://ui.adsabs.harvard.edu/abs/2024PhPl...31h2112P}
}

@ARTICLE{2025Patel,
       author = {{Patel}, Ritesh and {Niembro}, Tatiana and {Xie}, Xiaoyan and {Seaton}, Daniel B. and {Badman}, Samuel T. and {Roy}, Soumya and {Rivera}, Yeimy J. and {Reeves}, Katharine K. and {Stenborg}, Guillermo and {Hess}, Phillip and {West}, Matthew J. and {Feller}, Alex and {Hirzberger}, Johann and {Orozco Su{\'a}rez}, David and {Solanki}, Sami K. and {Strecker}, Hanna and {Valori}, Gherardo},
        title = "{Direct in situ observations of eruption-associated magnetic reconnection in the solar corona}",
      journal = {Nature Astronomy},
         year = 2025,
        month = aug,
          doi = {10.1038/s41550-025-02623-6},
       adsurl = {https://ui.adsabs.harvard.edu/abs/2025NatAs.tmp..161P}
}

@ARTICLE{2025PesceRollins,
       author = {{Pesce-Rollins}, Melissa and {MacKinnon}, Alexander and {Klein}, Karl-Ludwig and {Russell}, Alexander J.~B. and {Hudson}, Hugh and {Warmuth}, Alexander and {Wiegelmann}, Thomas and {Masson}, Sophie and {Parnell}, Clare and {Nitta}, Nariaki V. and {Omodei}, Nicola},
        title = "{Ion-rich Acceleration during an Eruptive Flux Rope Event in a Multiple Null-point Configuration}",
      journal = {\apj},
         year = 2025,
        month = aug,
       volume = {989},
       number = {2},
          eid = {148},
        pages = {148},
          doi = {10.3847/1538-4357/adeb7f},
archivePrefix = {arXiv},
       eprint = {2507.08469},
 primaryClass = {astro-ph.HE},
       adsurl = {https://ui.adsabs.harvard.edu/abs/2025ApJ...989..148P}
}

@ARTICLE{2018Phan,
       author = {{Phan}, T.~D. and {Eastwood}, J.~P. and {Shay}, M.~A. and {Drake}, J.~F. and {Sonnerup}, B.~U. {\"O}. and {Fujimoto}, M. and {Cassak}, P.~A. and {{\O}ieroset}, M. and {Burch}, J.~L. and {Torbert}, R.~B. and {Rager}, A.~C. and {Dorelli}, J.~C. and {Gershman}, D.~J. and {Pollock}, C. and {Pyakurel}, P.~S. and {Haggerty}, C.~C. and {Khotyaintsev}, Y. and {Lavraud}, B. and {Saito}, Y. and {Oka}, M. and {Ergun}, R.~E. and {Retino}, A. and {Le Contel}, O. and {Argall}, M.~R. and {Giles}, B.~L. and {Moore}, T.~E. and {Wilder}, F.~D. and {Strangeway}, R.~J. and {Russell}, C.~T. and {Lindqvist}, P.~A. and {Magnes}, W.},
        title = "{Electron magnetic reconnection without ion coupling in Earth's turbulent magnetosheath}",
      journal = {\nat},
         year = 2018,
        month = may,
       volume = {557},
       number = {7704},
        pages = {202-206},
          doi = {10.1038/s41586-018-0091-5},
       adsurl = {https://ui.adsabs.harvard.edu/abs/2018Natur.557..202P}
}

@ARTICLE{2013Phan,
       author = {{Phan}, T.~D. and {Shay}, M.~A. and {Gosling}, J.~T. and {Fujimoto}, M. and {Drake}, J.~F. and {Paschmann}, G. and {Oieroset}, M. and {Eastwood}, J.~P. and {Angelopoulos}, V.},
        title = "{Electron bulk heating in magnetic reconnection at Earth's magnetopause: Dependence on the inflow Alfv{\'e}n speed and magnetic shear}",
      journal = {\grl},
         year = 2013,
        month = sep,
       volume = {40},
       number = {17},
        pages = {4475-4480},
          doi = {10.1002/grl.50917},
       adsurl = {https://ui.adsabs.harvard.edu/abs/2013GeoRL..40.4475P}
}

@misc{AboutMMS,
  title = {{About MMS}},
  howpublished = {\url{https://lasp.colorado.edu/galaxy/spaces/mms/pages/2326576/About+MMS}},
  note = {Accessed: 2025-11-27}
}

\end{document}